\documentclass[%
 reprint,
superscriptaddress,
 amsmath,amssymb,
 aps,
prb,
]{revtex4-2}
\usepackage[utf8]{inputenc}
\usepackage[T1]{fontenc}

\usepackage{graphicx}
\usepackage{dcolumn}
\usepackage{bm}
\usepackage{color}
\usepackage{amsmath}
\usepackage{graphics}
\usepackage{subfigure}
\usepackage{pdftexcmds}
\usepackage{ifpdf}
\usepackage{amssymb}
\usepackage{braket}
\usepackage{dsfont}

\pdfoutput=1

\newcommand{\iu}{\mathrm{i}}
\newcommand{\CrNbS}{$\mathrm{CrNb}_3\mathrm{S}_6$}
\newcommand{\hn}{\boldsymbol{n}}
\newcommand{\hx}{\boldsymbol{x}}
\newcommand{\hy}{\boldsymbol{y}}
\newcommand{\hz}{\boldsymbol{z}}
\newcommand{\hu}{\boldsymbol{u}}

\newcommand{\he}{\boldsymbol{e}}

\newcommand{\vB}{\boldsymbol{B}}

\renewcommand{\vr}{\boldsymbol{r}}
\newcommand{\vtau}{\boldsymbol{\tau}}
\newcommand{\vj}{\boldsymbol{j}}
\newcommand{\vk}{\boldsymbol{k}}

\newcommand{\hc}{h_{\mathrm{c}}}

\newcommand{\pmin}{p_{\mathrm{min}}}
\newcommand{\pmax}{p_{\mathrm{max}}}

\begin{document}

\title{The continuum of metastable conical states of monoaxial chiral helimagnets}

\author{V. Laliena}
\affiliation{
Department of Applied Mathematics and
Institute of Mathematics and Applications (IUMA), University of Zaragoza
C/ Mar\'ia de Luna 3, 50018 Zaragoza, Spain
}

\author{S. A. Osorio}
\affiliation{ 
Instituto de Nanociencia y Nanotecnolog\'ia (CNEA-CONICET), Nodo Bariloche, Av. Bustillo 9500 (R8402AGP), S. C. de Bariloche, R\'io Negro, Argentina
}
\affiliation{
Gerencia de F\'isica,  Centro At\'omico Bariloche, Av. Bustillo 9500 (R8402AGP), S. C. de Bariloche, R\'io Negro, Argentina
}

\author{D. Bazo}
\affiliation{
Aragon Nanoscience and Materials Institute (CSIC-University of Zaragoza) and Condensed Matter Physics Department, University of Zaragoza, C/ Pedro Cerbuna 12, 50009 Zaragoza, Spain
}
\affiliation{
Department of Applied Mathematics and
Institute of Mathematics and Applications (IUMA), University of Zaragoza
C/ Mar\'ia de Luna 3, 50018 Zaragoza, Spain
}

\author{S. Bustingorry}
\affiliation{ 
Instituto de Nanociencia y Nanotecnolog\'ia (CNEA-CONICET), Nodo Bariloche, Av. Bustillo 9500 (R8402AGP), S. C. de Bariloche, R\'io Negro, Argentina
}
\affiliation{
Gerencia de F\'isica,  Centro At\'omico Bariloche, Av. Bustillo 9500 (R8402AGP), S. C. de Bariloche, R\'io Negro, Argentina
}
\affiliation{
Aragon Nanoscience and Materials Institute (CSIC-University of Zaragoza) and Condensed Matter Physics Department, University of Zaragoza, C/ Pedro Cerbuna 12, 50009 Zaragoza, Spain
}

\author{J. Campo}
\affiliation{
Aragon Nanoscience and Materials Institute (CSIC-University of Zaragoza) and Condensed Matter Physics Department, University of Zaragoza, C/ Pedro Cerbuna 12, 50009 Zaragoza, Spain
}

\date{\today}

\begin{abstract}

At low temperature and zero applied magnetic field, besides the equilibrium helical state, monoaxial chiral helimagnets have a continuum of helical states differing by the wave number of the modulation. The wave number of these states in units of the equilibrium state wave number is denoted here by $p$, and accordingly the corresponding states are called the $p$-states.
In this work we study in detail the metastability of the $p$-states. The application of an external magnetic field in the direction of the chiral axis has a double effect: on one hand, it introduces a conical deformation of the $p$-states, and on the other hand it destabilizes some of them,  shrinking the range of $p$ in which the $p$-states are metastable. If a polarized current is applied along the chiral axis, the $p$-states reach a steady moving state with a constant velocity proportional to the current intensity. 
Besides this dynamical effect, the polarized current also induces a conical deformation and reduces the range of stability of the $p$-states.
The stability diagram in the plane applied field - applied current intensity has interesting features that, among other things, permit the manipulation of $p$-states by a combination of applied fields and currents. These features can be exploited to devise processes to switch between $p$-states. In particular there are $p$-states with negative $p$, opening the possibility to helicity switching.
The theoretical feasibility of such processes, crucial from the point of view of applications, is shown by micromagnetic simulations.
Analogous $p$-states exists in cubic chiral helimagnets and therefore similar effects are expected in those systems.

\end{abstract}

\maketitle

\section{Introduction}

Noncollinear magnetic textures such as magnetic helices, domain walls, vortices, or skyrmions are very promising for spintronic applications due to the possibility to control them using different external stimuli, like magnetic fields or polarized electric currents \cite{Berger84,Freitas85,Chappert07,Parkin08,Nagaosa13,Back20}.
To be useful, these magnetic textures have to be (meta)stable in some part of the relevant parameter space.
Noncollinear magnetic textures appear, in particular, as equilibrium states at low temperature in chiral magnets, which are
characterized by the presence of a sizable Dzyaloshinskii-Moriya interaction (DMI). 
The most studied systems of this kind are cubic chiral helimagnets and films with interfacial DMI, which 
host skyrmions and skyrmion lattices \cite{Muehlbauer09,Yu10,Yu11,Nagaosa13,Back20}.
Monoaxial chiral helimagnets, in which the DMI propagates only along a single direction (the chiral axis), have received comparatively less attention.
Besides the archetypal \CrNbS, other known monoaxial chiral helimagnets are MnNb$_3$S$_6$, CrTa$_3$S$_6$, CuB$_2$O$_4$, CuCsCl$_3$, Yb(Ni$_{1-x}$Cu$_x$)$_3$Al$_9$, and Ba$_2$CuGe$_2$O$_7$~\cite{Moriya82,Kousaka16,Roessli01,Adachi80,Ohara14,Matsumura17,Zheludev97,Karna2019, Karna2021}.

Not surprisingly, monoaxial chiral helimagnets present also a strong uniaxial magnetic anisotropy (UMA) 
along the chiral axis, which is of easy-plane type in \CrNbS. 
The competition of the exchange interaction, the DMI, the UMA and the 
applied field determines the equilibrium state at low enough temperature, where  thermal fluctuations are only a minor effect. At zero external field the equilibrium state is a magnetic helix 
with wave vector along the chiral axis and wave number determined by the competition of the 
exchange interaction and the DMI. If a low enough external field is applied in a direction perpendicular to the chiral axis the equilibrium state is a Chiral Soliton Lattice (CSL)\cite{Dzyal64,Izyumov84,Kishine05,Togawa12}. If instead the magnetic field is applied in the direction of the chiral axis, the equilibrium state is a conical state~ \cite{Miyadai83,Izyumov84,Ghimire13,Chapman14,Laliena16a,Laliena17a}. These two magnetic textures, the CSL and the conical state, are of different nature: the CSL is solitonic while the conical state is helical. If the field direction is neither perpendicular nor parallel, the equilibrium state is a one-dimensional modulated texture which connects smoothly the two limiting cases as the direction of the magnetic field is varied from perpendicular to parallel to the chiral axis~\cite{Laliena16a}. In all cases, for sufficiently large magnetic fields the equilibrium state is the forced ferromagnetic state (FFM), which has a uniform magnetization pointing in the direction of the external field. The different nature of the CSL and the conical states is manifested in the transition to the FFM state:
in the former case it is of nucleation type and in the latter of instability type, in de Gennes's terminology \cite{DeGennes75}. These two different kinds of phase boundaries are separated by tricritical points in the temperature - applied field phase diagram \cite{Laliena16b,Laliena17a}.
The phase diagram of monoaxial chiral helimagnets and the nature of the phase boundaries in the  temperature-applied magnetic field space, determined experimentally by several groups~\cite{Ghimire13,Tsuruta16,Yonemura17,Clements17,Clements18}, agree well with these theoretical predictions.

It was shown in \cite{Laliena18a} that, at low temperature, besides the conical equilibrium state a continuum of conical states differing by the wave number and the magnetization component along the chiral axis are local minima of the energy functional of monoaxial chiral helimagnets with external magnetic field applied along the chiral axis. Similar local minima of the energy are present in cubic chiral helimagnets \cite{Laliena17b}. 
These conical states, called here $p$-states, include states with helicity
opposite to the helicity of the equilibrium state, which, although energetically disfavoured by the 
DMI, remain as metastable states in some range of the applied magnetic field.
It is also remarkable that among these continuum of metastable states there are some which are 
ferromagnetic, with the uniform magnetization pointing to a direction determined by the competition between the UMA and the applied field.

In this work we analyze in detail the properties of these $p$-states of monoaxial chiral helimagnets, clarifying their role as metastable states and studying their behavior under the action of polarized electric currents. One conclusion of this analysis is the possibility of switching between metastable conical states with different wave vectors, including the possibility of helicity reversing. This is clearly of great interest for applications. Indeeed, it has been argued recently that controlled switching among magnetic states with opposite helicity might be used for memory applications~\cite{Jiang2020}.

\section{A continuum of conical states \label{sec:conical}}

Consider a monoaxial chiral helimagnet, such as \CrNbS, with chiral axis along $\hz$ (we shall use $\hx, \hy, \hz$ as the orthonormal vector triad in space). 
At low enough temperature the local magnetization is given by $M_\mathrm{S}\hn$, where $\hn$ is a unit vector field that describes the magnetization direction at each point of the material and the constant $M_\mathrm{S}$ is the saturation magnetization.
The magnetic energy is given by 
the functional $E[\hn] = \int d^3r \, e(\vr)$, with 
\begin{equation}
  e(\vr) = A \sum_i (\partial_i\hn)^2-D\hz\cdot(\hn\times\partial_z\hn)
  - K(\hz\cdot\hn)^2-M_\mathrm{S}B\hz\cdot\hn. \label{eq:E}
\end{equation}
In the above equation the index $i$ runs over $\{x,y,z\}$, $A$, $D$, and $K$ stand for the exchange stiffnes constant, and the DMI and UMA strength constants, respectively, and $B\hz$ is the applied magnetic field. We consider $K < 0$ to have an easy plane perpendicular to $\hz$. The DMI acts only along the $\hz$ axis, defining the chiral axis (notice that the external field is applied along the chiral axis). The sign of $D$ is reversed if we reverse the direction of the $\hz$ axis, so that, with no loss of generality, we take $D>0$.
It is convenient to introduce the parameters
\begin{equation}
q_0=\frac{D}{2A}, \quad \kappa=\frac{4AK}{D^2}, \quad h=\frac{2AM_\mathrm{S}}{D^2}B.
  \label{eq:params}
\end{equation}
Notice that $q_0$ has the dimensions of inverse length while $\kappa$ and $h$ are dimensionless. 
We do not include explicitly in the energy the magnetostatic energy, whose effect in an infinite system in which the magnetization depends only on $z$ (as it is in this work) is completely absorbed in the UMA \cite{Hubert08}.

The dynamics of $\hn$ obeys the Landau-Lifschitz-Gilbert (LLG) equation
\begin{equation}
\partial_t\hn = \gamma\vB_{\mathrm{eff}}\times\hn + \alpha\hn\times\partial_t\hn +\vtau, \label{eq:LLG}
\end{equation}
where $\alpha$ and $\gamma$ are the Gilbert damping parameter and the gyromagnetic constant, respectively, and $\vtau$ stands for some applied 
nonconservative torque not included in the energy (\ref{eq:E}).
The effective field acting on $\hn$ is given by 
\begin{equation}
\vB_{\mathrm{eff}} \!= \!\frac{2A}{M_\mathrm{S}} \!\Big( \nabla^2\hn - 2q_0\hz\times\partial_z\hn+ q_0^2\kappa(\hz\cdot\hn)\hz + q_0^2 h\hz \Big).
\end{equation}

\begin{figure}[t!]
\centering
\includegraphics[width=\linewidth]{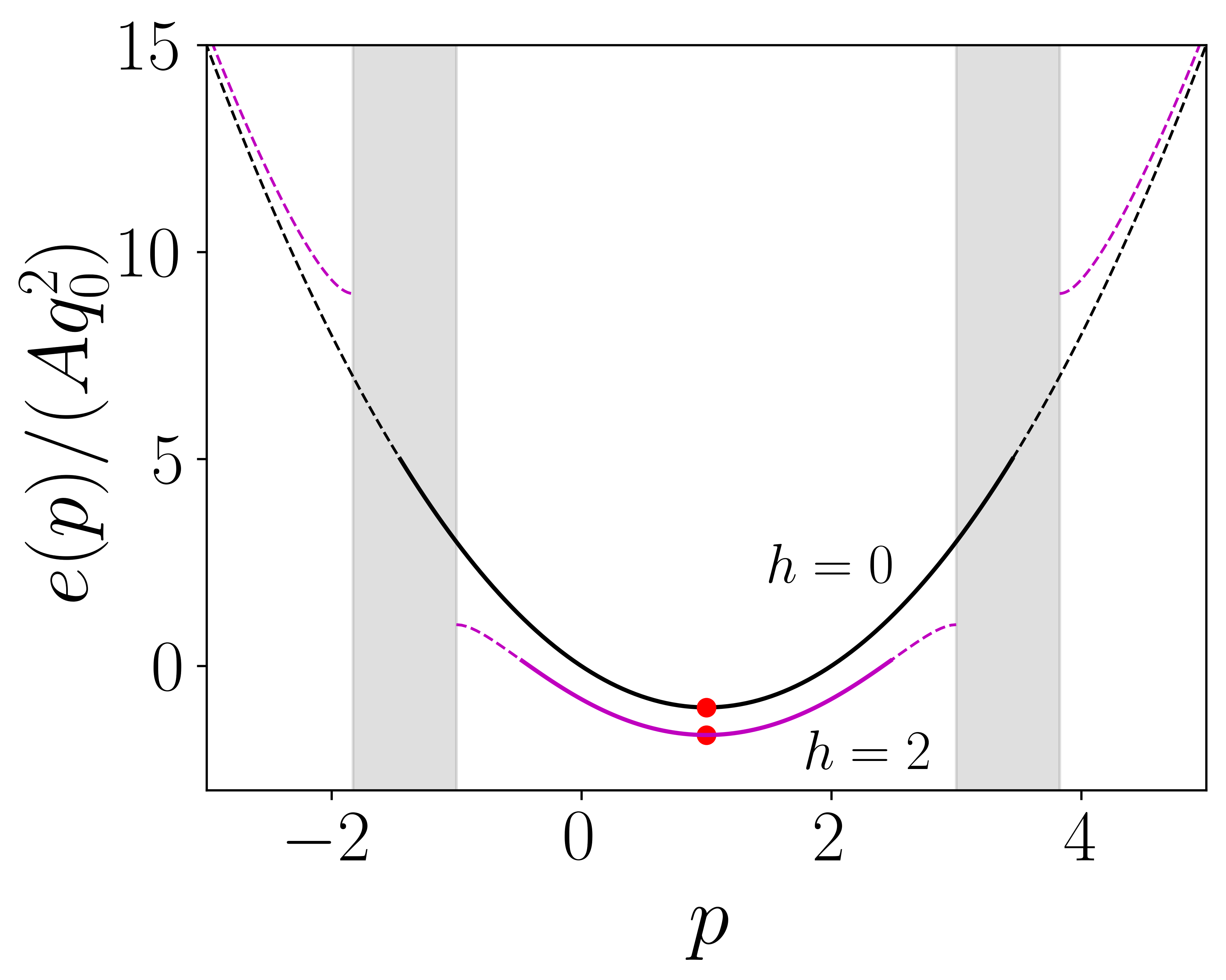}
\caption{Energy density for $h=0$ and $h=2<\hc$, as a function of $p=q/q_0$. The minimum value corresponds always to the equilibrium state with $p=1$.
States indicated with continuous lines are stable against localized deformations, while dashed lines indicate unstable states, as shown in Sec.~\ref{sec:Stability-conical}. The grey regions indicates the gap in $p$ values where there are no states satisfying $|\cos \theta_p| \leq 1$.
\label{fig:eph}}
\end{figure}

In absence of external torque $\vtau$ the equilibrium states are solutions of the static equation $\vB_{\mathrm{eff}} = \lambda\hn$, where $\lambda$ is a Lagrange multiplier enforcing the constraint $\hn^2=1$.
For $h\geq\hc$, where $\hc=1-\kappa > 1$ is the critical field, the equilibrium state is the homogeneous FFM state, with the magnetization pointing along the $\hz$ direction: $\hn=\hz$. For $h<h_c$ the static equation admits solutions which are modulated states with the form of a conical helix propagating along the chiral axis. 
With the parametrization
\begin{equation}
\hn = \sin\theta\cos\varphi\,\hx + \sin\theta\sin\varphi\,\hy + \cos\theta\,\hz,
\label{eq:hn}
\end{equation}
these modulated states are given by \cite{Laliena18b}
\begin{equation}
\label{eq:cosh0}
    \cos\theta_p = \frac{h}{\hc-(p-1)^2}, \qquad \varphi_p(z)=p q_0 z,
\end{equation}
where $p$ is the wave number in units of $q_0$. It is convenient to label these states by $p$, writing $\theta_p$, $\varphi_p$ and $\hn_p$. For the sake of brevity, these states 
will be referred to as the $p$-states, i.e. a $p$-state is a conical state with wave number 
$q = p q_0$. Since $|\cos \theta_p| \leq 1$, the range of $p$ is limited to
\begin{equation}
    1-\sqrt{\hc-|h|} \leq p \leq 1+\sqrt{\hc-|h|}. \label{eq:existence}
\end{equation}
Notice that $\hc>1$ because we consider easy-plane anisotropy. Hence, for $|h|$ small enough $p$ can be negative. These $p$-states have helicity against the DMI. In a range of $h$ there is also a state with $p=0$, which is a ferromagnetic state with the magnetization component along the chiral axis given by $n_z=h/(\hc-1)$. We shall comment on these rather unexpected states in Sec.~\ref{sec:features}.

The energy density of the $p$-states is given by
\begin{equation}
\label{eq:edep}
    e(p) = A q_0^2 \left[ (p-1)^2-1- \frac{h^2}{\hc-(p-1)^2} \right].
\end{equation}
The minimum of the energy corresponds to $p = 1$ for all $|h| \leq \hc$, what means that the equilibrium states are those with $p=1$ (wave number $q_0$).
It is shown in Sec. \ref{sec:Stability-conical} that for $|h| < \hc$ there exists a range of $p$ around $p=1$ in which $p$-states are metastable. This implies that these $p$-states are local minima of the energy functional and, therefore, the small perturbations around them are damped as they evolve according to the LLG equation.

The energy density of the $p$-states as a function of $p$ is displayed in Fig.~\ref{fig:eph} for $h=0$ and $h=2<\hc$, where $\hc = 6$ approxiamately corresponds to \CrNbS. The state with minimum energy corresponds always to $p=1$ (red dots). The metastable $p$-states are located in a finite range around $p=1$ signaled by the continuous lines. Outside this range the $p$-states are unstable (dashed lines). 
Notice that for $h\neq 0$ there is a gap in $p$ values for which there are no states satisfying $|\cos \theta_p| \leq 1$.

It is remarkable that, in spite of what the form of Fig.~\ref{fig:eph} may suggest, 
states with $p \neq 1$ are metastable since the value of $p$ cannot be changed by small perturbations.
Indeed, consider a small variation $\delta p$ of $p$. A straightforward computation shows that for $\delta p\to 0$
\begin{equation}
    \hn_{p+\delta p} - \hn_{p} \sim 2\sin\theta_p\sin\left(\frac{\delta p q_0 z}{2}\right) \hu(z),
\end{equation}
where
\begin{equation}
    \hu(z) = -\sin\big((p+\delta p/2)q_0z\big) \hx + \cos\big((p+\delta p/2)q_0z\big) \hy 
\end{equation}
is a unit vector. This means that a small change of $p$ cannot be considered a small perturbation of $\hn_p$, since $|\hn_{p+\delta p}-\hn_p|$ is not small for
$\delta p q_0 z$ close to $\pi$. This may be clearer in 
a bounded system, of length $R$, with periodic boundary conditions: then the minimum $\delta p$ is $2\pi/q_0R$ and $0\leq z \leq R$, so that for this minimum $\delta p$ we have $\delta p q_0 z = \pi$ if $z=R/2$. 
Summarizing, the situation is the following: 1) the metastability of a state is related to the behaviour of its energy under small perturbations; 2) a change of $p$, however small, is not a small perturbation of the $p$-state; 3) it is incorrect to infer from Fig.~\ref{fig:eph} that the $p$-states are not metastable.

The discussion of the previous paragraph implies that although the energy density of the $p$-states corresponding to $p$ and $p+\delta p$ is close, if the stability ellipses of $p$ and $p+\delta p$ enclose the point $(h,\Gamma=0)$ these $p$-states are separated by energy barriers in the whole configuration space of $\hn$, for this value of $h$.
How long is the life time of the metastable $p$-states depends on these energy barriers. This is a question that cannot be tackled with the methods used in this work.
In any case, we expect that the lifetime will increase by decreasing the temperature, a question that deserves further study.

\section{Steady motion of the conical states under the action of a polarized current}

In this section we study the response of the $p$-states to a polarized electric current applied along the chiral axis. If the current density is $\vj = -j\hz$, the magnetic torque delivered by the current is given by
\begin{equation}
\vtau = -jb_j\big(\partial_z\hn  - \beta \hn\times\partial_z\hn\big),
\label{eq:torque}
\end{equation}
with $b_j=P \mu_\mathrm{B}/(|e|M_\mathrm{s})$, where $P$ is the polarization degree of the current, $e$ is the electron charge, and $\mu_B$ is the Bohr magneton \cite{Zhang04}.
The first term is the reactive (adiabatic) torque and the second term the dissipative (non-adiabatic) torque, whose strength is controlled by the nonadiabaticity coefficient $\beta$~\cite{Thiaville05}.


We start by seeking for steady solutions of the LLG equation (\ref{eq:LLG}) which have the form of a state that moves rigidly with constant velocity, $v$,
along the $\hz$ direction. The general steady solution is characterized by two functions, $\theta(w)$ and $\varphi(w)$, of the variable $w=q_0(z-vt)$.
Inserting this \textit{ansatz} in the LLG equations we obtain the steady motion equations, which can be cast to the form
\begin{gather}
 \theta^{\prime\prime} - (\varphi^\prime-1)^2\sin\theta\cos\theta + (\hc\cos\theta-h)\sin\theta + \nonumber \\
\Omega\theta^\prime-\Gamma\sin\theta\varphi^\prime =0,
\label{eq:theta} \\[6pt]       
\sin\theta\varphi^{\prime\prime} +2\cos\theta\theta^\prime(\varphi^\prime-1) + \Gamma\theta^\prime + \Omega\sin\theta\varphi^\prime=0,
\label{eq:phi}
\end{gather}
with the primes standing for derivatives with respect to $w$ and
\begin{eqnarray}
\Omega &=& \frac{\alpha q_0}{\omega_0}\left(v-\frac{\beta}{\alpha}b_jj\right), \label{eq:Omega} \\
\Gamma &=& \frac{q_0}{\omega_0} \left(v - b_jj\right), \label{eq:Gamma}
\end{eqnarray}
where the quantity $\omega_0=2 \gamma q_0^2 A /M_\mathrm{S}$
has the dimensions of a frequency.
Notice that the spin transfer torque, the Gilbert damping, the nonadiabaticity coefficient, and the steady velocity enter the equations of steady motion only through the parameters $\Omega$ and $\Gamma$.

The solutions of Eqs. (\ref{eq:theta}) and (\ref{eq:phi}) with constant $\theta=\theta_p$ and $\varphi^\prime=p$ correspond to steady moving $p$-states.
In this case Eq. (\ref{eq:theta}) is satisfied if
\begin{equation}
\cos\theta_p = \frac{h+p \Gamma}{\hc-(p-1)^2}. \label{eq:ct}
\end{equation}
This steady moving $p$-state exists only if $|h+p\Gamma|\leq \hc-(p-1)^2$. 
The stability of these solutions is analyzed in Sec.~\ref{sec:Stability-conical}.

To have a solution with constant $\theta=\theta_p$ and $\varphi^\prime=p$ Eq.~(\ref{eq:phi}) requires $\Omega=0$, what provides the relation between the steady velocity and the intensity of the applied current,
\begin{equation}
v = \frac{\beta}{\alpha}b_jj, \label{eq:v}
\end{equation}
and thus $\Gamma$ becomes proportional to the current density,
\begin{equation}
  \Gamma = \frac{(\beta-\alpha)q_0}{\alpha\omega_0}b_jj. \label{eq:jGamma}
\end{equation}
We see that the steady state velocity increases linearly with the current density, with a mobility $m=(\beta/\alpha)b_j$ which is independent of the system parameters $\kappa$ and $h$. 
The same behavior occurs for domain walls \cite{Thiaville05}, for 360$^o$ domain walls \cite{Mascaro10,Jin16}, and for the isolated solitons and the chiral soliton lattice of monoaxial chiral helimagnets \cite{Laliena20,Osorio2022,Kishine10}.
Therefore this relation between steady velocity and applied current density seems to be a universal feature of the response of one dimensional magnetic modulated states to polarized currents.

Equation~(\ref{eq:v}) implies that $v=0$ if $\beta=0$, so that the steady moving solution is actually static if there is no dissipative torque. In this case, after applying the current the system reaches a different equilibrium state, a static $p$-state with cone angle given by equation (\ref{eq:ct}).
Notice also that the case $\beta=\alpha$ is special, since then $\Omega=0$ and $\Gamma=0$, and therefore Eqs. (\ref{eq:theta}) and (\ref{eq:phi}) are independent of the applied current. This implies that in this case the $p$-state is rigidly dragged by the current, with velocity $v=b_jj$, keeping the cone angle equal to its static value.

\section{Stability of the magnetic states}
\label{sec:stability}

In this section we analyze the stability of magnetic states against small perturbations. The section is divided into three subsections: one dealing with the stability of the $p$-states, another one devoted to the the stability of the FFM state, and the last one in which we discuss the main features of the stability diagram.

\subsection{Stability of the conical states}
\label{sec:Stability-conical}

We analyze here the stability of the generic steady moving $p$-state obtained for given $h$ and $\Gamma$. The static $p$-states discussed in Sec.~\ref{sec:conical} are the particular cases $\Gamma=0$ of this general analysis. Here a $p$-state is a steady moving state if $\Gamma\neq 0$ and a static state if $\Gamma=0$.

Let $\hn_p$ be the (unitary) magnetization field of the steady moving $p$-state, with $\theta_p$
described by (\ref{eq:ct}) and $\varphi_p=pq_0(z-vt)$, with $v$ given by (\ref{eq:v}). A small perturbation of $\hn_p$ is given by two fields,
$\xi_1$ and $\xi_2$, which depend on the three coordinates $x$, $y$, $z$, and on time $t$, so that, for small enough $\xi_1$ and $\xi_2$,
the perturbed magnetization is given by
\begin{equation}
  \hn = \sqrt{1-\xi_1^2+\xi_2^2}\,\hn_p + \xi_1\,\he_1+\xi_2\,\he_2, \label{eq:n2}
\end{equation}
where $\{\he_1,\he_2,\hn_p\}$ form a right-handed orthonormal triad. We take
\begin{eqnarray}
  \he_1&=&\cos\theta_p\cos\varphi_p\,\hx + \cos\theta_p\sin\varphi_p\,\hy - \sin\theta_p\,\hz, \qquad \\
  \he_2&=&-\sin\varphi_p\,\hx+\cos\varphi_p\,\hy.
\end{eqnarray}
We require that, for fixed $t$, the fields $\xi_1$ and $\xi_2$ be square integrable functions of $(x,y,z)$, to ensure that the energy of the perturbation is finite.

The perturbed magnetization has to be a solution of the LLG equation. 
Inserting Eq.~(\ref{eq:n2}) into (\ref{eq:LLG}) we obtain the equations
for the dynamics of the perturbation $\xi=(\xi_1,\xi_2)^T$. Expanding in powers of $\xi_1$ and $\xi_2$, we have to linear order
\begin{equation}
\partial_t\xi = \frac{\omega_0}{(1+\alpha^2)q_0^2} \, \mathcal{D}\xi, \label{eq:LLGlin2}
\end{equation}
where $\mathcal{D}$ is a $2\times 2$ operator matrix whose matrix elements are the linear differential operators
\begin{eqnarray}
  &&\mathcal{D}_{11}=\alpha(\nabla^2-a)+\big[\Delta-(1+\alpha\beta)b\big]\partial_z, \label{eq:D11p} \\[6pt]
  &&\mathcal{D}_{12} = \nabla^2-\big[\alpha\Delta+(\beta-\alpha)b\big]\partial_z, \label{eq:D12p} \\[6pt]
  &&\mathcal{D}_{21}=-\nabla^2+a+\big[\alpha\Delta+(\beta-\alpha)b\big]\partial_z, \label{eq:D21p} \\[6pt]
  &&\mathcal{D}_{22} = \alpha\nabla^2+\big[\Delta-(1+\alpha\beta)b\big]\partial_z, \label{eq:D22p}
\end{eqnarray}
with
\begin{gather}
a=q_0^2\,\big(\hc-(p-1)^2\big)\sin^2\theta_p,  \label{eq:ap} \\[8pt]
\Delta=q_02(p-1)\cos\theta_p, \qquad
b=\frac{q_0\alpha}{\beta-\alpha}\Gamma,  \label{eq:dbp}
\end{gather}
where we assumed $\alpha\neq\beta$. The case $\alpha=\beta$ is special, as we said before, since then $\Gamma=0$ for any value of the applied current. In this case
$b=q_0^2b_jj/\omega_0$.

Stability requires that the spectrum of $\mathcal{D}$ lies in the complex half-plane with non negative real part.
Since $\mathcal{D}_{ij}$ are linear differential operators with constant coefficients, the spectrum of $\mathcal{D}$ can be readily obtained by Fourier transform.
Details on the calculations leading to the stability conditions are given in Appendix \ref{app:conical}. Here we collect the conclusions. A necessary condition for stability is $a\geq 0$, 
what gives the following bounds for the $p$ values of stable $p$-states:
\begin{equation}
1-\sqrt{\hc} \leq p \leq 1+\sqrt{\hc}.
\label{eq:pbounds}
\end{equation}
It is shown in Appendix~\ref{app:conical} that, 
having $p$ within these bounds, the $p$-state is stable only in the region of the $(h, \Gamma)$ plane enclosed by the ellipse of equation
\begin{equation}
A(p)\Gamma^2+2B(p)\Gamma h+C(p) h^2 = D(p), \label{eq:ellipse}
\end{equation}
where the functions $A(p)$, $B(p)$, $C(p)$, and $D(p)$ are independent of $h$ and $\Gamma$, 
and are given in Appendix~\ref{app:conical}.
The stability ellipses of the $p$-states are centered at $(0, 0)$ and have the principal axes rotated with respect to the coordinate axes. The amount of rotation depends on $p$.

The stability of the static $p$-states discussed in Sec.~\ref{sec:conical} is obtained by setting $\Gamma=0$ in this general approach. Thus, the static $p$-state is stable 
in the range of $h$ determined by the intersection of its stability ellipse with the $\Gamma=0$ axis.

\begin{figure}[t!]
\centering
\includegraphics[width=\linewidth]{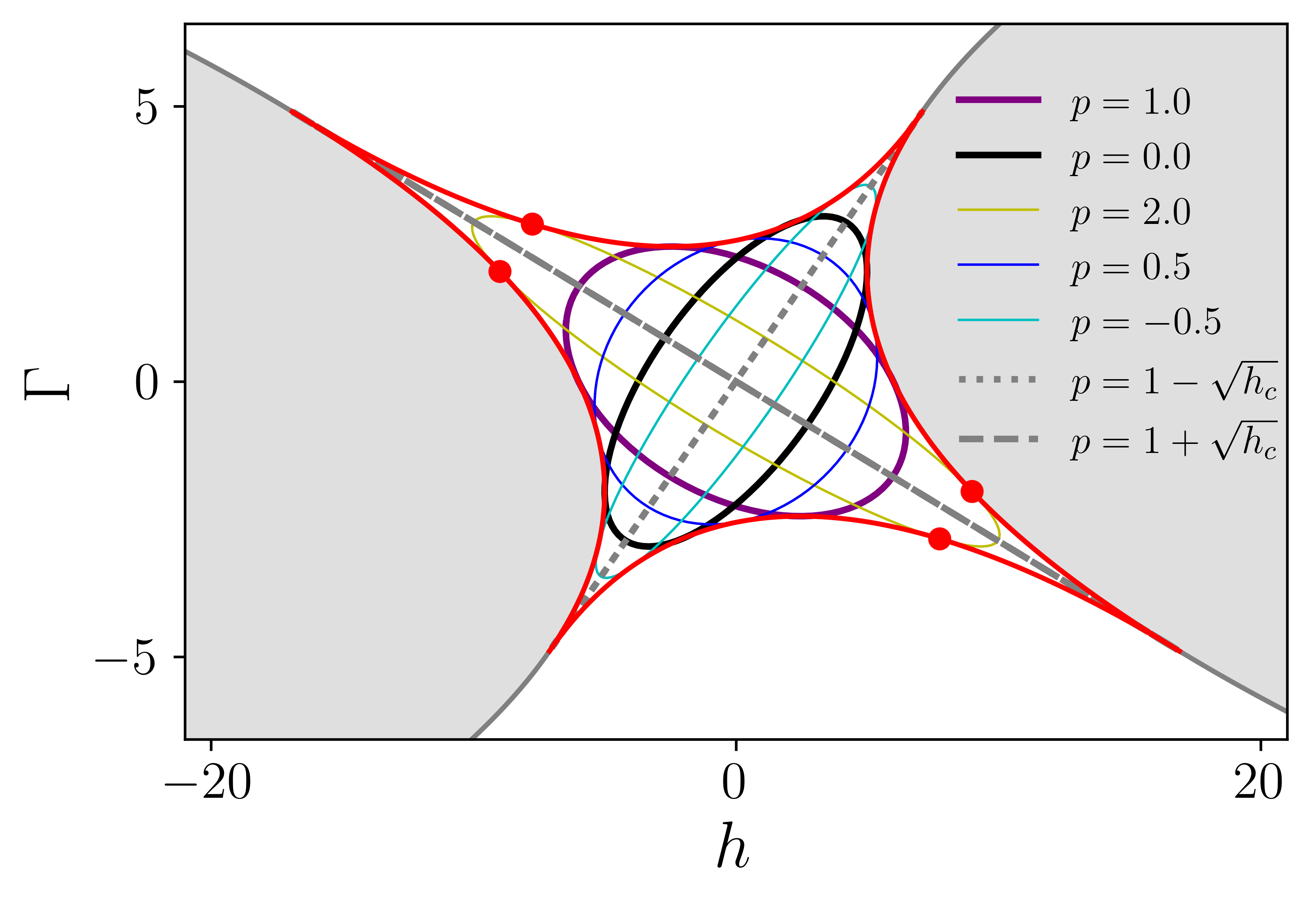}
\caption{
  Stability diagram in the plane $(h,\Gamma)$ for $\hc=6$. Steady moving $p$-states are stable inside the region bounded by the red line. The FFM states are only stable within shaded regions, which are unbounded. Red dots corresponds to the stability limit of states with $p=2$ lying in the boundary of the stability diagram, as those shown in Fig.~\ref{fig:pbound}.
\label{fig:phd}}
\end{figure}

Figure~\ref{fig:phd} displays the stability ellipses for several values of $p$ in the $(h,\Gamma)$ plane, for $\hc=6$, which approximately corresponds to \CrNbS.
For each $p$ value, the $p$-state is metastable for $(h,\Gamma)$ inside the corresponding ellipse, 
and unstable outside it.

The region of the $(h,\Gamma)$ plane in which there exists some stable steady moving $p$-state is
bounded by the envelope of the one-parametric family of ellipses (parametrized by $p$) given by Eq. (\ref{eq:ellipse}).
The envelope can be readily found and it has four branches given by
\begin{equation}
\label{eq:envelope}
\left.
\begin{array}{l}
  \Gamma = \sigma(h)2\left[(1\pm\sqrt{|h|-(\hc-1)}\right], \\[8pt]
  \Gamma = -h\pm 2\sqrt{\hc}\left[\sqrt{\hc(\hc-1)-\sqrt{\hc}h}-\hc\right],
\end{array}
\right.
\end{equation}
where $\sigma(h)$ is the sign function: $\sigma(h)=1$ if $h\geq 0$ and $\sigma(h)=-1$ if $h<0$.
The parametric equations of the envelop are given in Appendix \ref{app:conical}, Eqs.~\eqref{eq:env1}-\eqref{eq:env4}.
We call the region enclosed by this envelope the \textit{stability region of conical states}.
No modulated state is stable outside this region.

\begin{figure}[t!]
\centering
\includegraphics[width=\linewidth]{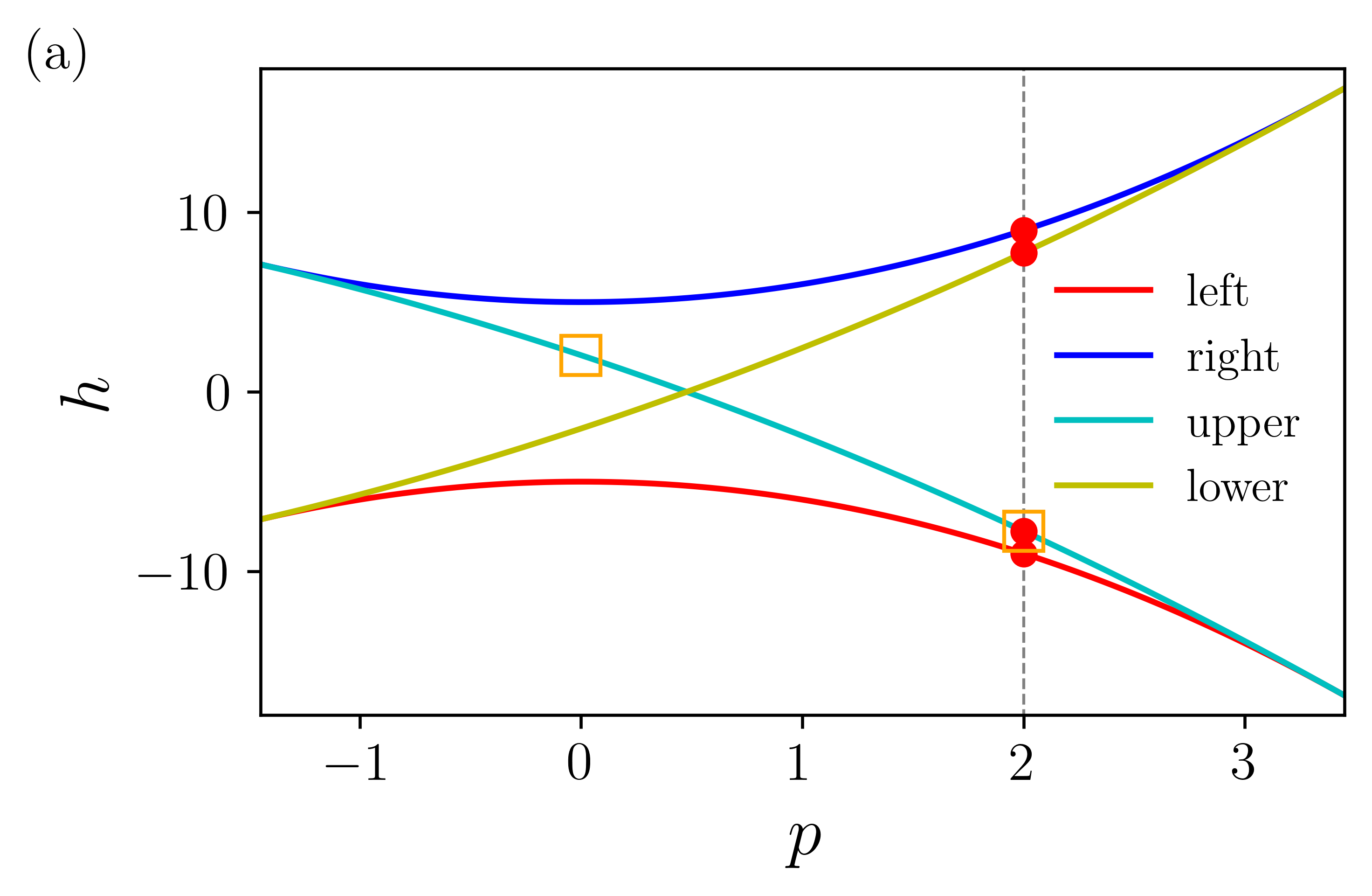}
\includegraphics[width=\linewidth]{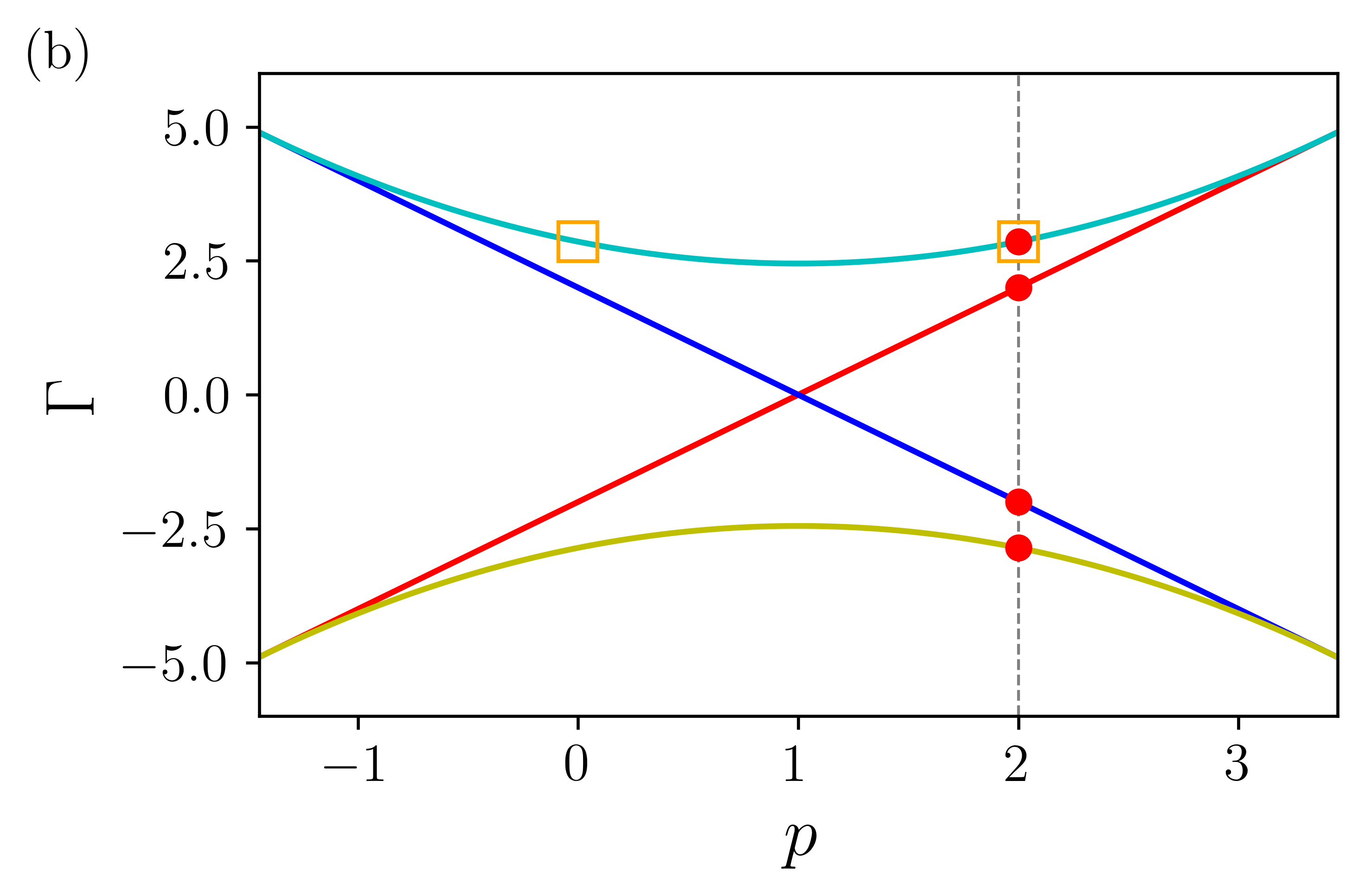}
\caption{
The values of (a) $h$ and (b) $\Gamma$ as a function of $p$ along the different branches of the envelope of the stability region defined by Eqs.~\eqref{eq:envelope}. For a given value of $p$ the four branches (left, right, upper, lower) are presented, corresponding to those shown in Fig.~\ref{fig:phd}.
Red dots correspond to those shown in Fig.~\ref{fig:phd}. Square symbols indicate $(h,\Gamma)$ points for which states with $p=0$ and $p=2$ are obtained in Sec.~\ref{sec:numerics}.
\label{fig:pbound}}
\end{figure}

The four branches of the envelope in the case $\hc=6$ are shown in red in Fig.~\ref{fig:phd}. 
Along each branch $p$ changes continuously within its bounds, from $1-\sqrt{\hc}$ to $1+\sqrt{\hc}$.
Each ellipse, determined by a given value of $p$, is tangent to the envelope at four points, one for each branch.
These four points, which depend on $p$, define the four pairs of functions shown in Fig.~\ref{fig:pbound}. The red points in Figs.~\ref{fig:phd} and \ref{fig:pbound} correspond to $p=2$. The detailed features of the stability diagram will be further discussed in Sec.~\ref{sec:features}.

\subsection{Stability of the forced ferromagnetic state}
\label{sec:Stability-FFM}

The FFM state is the magnetic texture with a uniform magnetization aligned with the applied field, which in our case points to the direction of the chiral axis $z$. Hence, the FFM is given by $\hn=\hz$ if $h\geq 0$ and $\hn=-\hz$ if $h<0$. It is the equilibrium state if $|h|>\hc$.
The FFM state is insensitive to an applied current, since the torque (\ref{eq:torque}) vanishes for uniform magnetization. However, it is destabilized by a sufficiently intense current. 
In this section we discuss the stability diagram of the FFM state in the applied field - applied current plane.

The perturbed FFM state has the form
\begin{equation}
  \hn = \sqrt{1-\xi_1^2-\xi_2^2}\,\sigma(h)\,\hz + \xi_1\,\sigma(h)\,\hx + \xi_2\,\hy.
\end{equation}
where $\sigma(h)=1$ if $h\geq 0$ and $\sigma(h)=-1$ if $h<0$.
The dynamics of the perturbation is obtained by inserting the above expression in the LLG equation  (\ref{eq:LLG}).
Again, the linearized LLG equation is given by a linear differential operator with constant coefficients whose spectrum is obtained by Fourier transform. The stability of the FFM state requires that the spectrum lies in the complex plane with non positive real part. The details of the calculations are given in Appendix~\ref{app:FFM}. The stability condition leads to the inequality
\begin{equation}
\Gamma^2-4\sigma(h)\Gamma+4(\hc-|h|)\leq 0, \label{eq:FFMstab}
\end{equation}
where $\Gamma$ is related to the current intensity, $j$, by Eq. \eqref{eq:jGamma}.
The FFM state is stable in the region of the $(h,\Gamma)$ plane in which the above inequality holds. 

Inequality \eqref{eq:FFMstab} holds if and only if the two roots in $\Gamma$ of the left hand side of the inequality are real, and $\Gamma$ is between the two roots. Then $|h|>\hc-1$ and
\begin{equation}
2\big(1- \zeta(h) \big)\leq \sigma(h)\Gamma \leq 2\big(1+ \zeta(h) \big), \label{eq:FFMstability}
\end{equation}
with $\zeta(h) = \sqrt{1+|h|-\hc}$.
The above inequalities determine the region of stability of the FFM state in the $(h,\Gamma)$ plane, which is displayed in Fig.~\ref{fig:phd} for $\hc=6$. 

It is remarkable that the boundary of the stability region of the FFM state coincides exactly with the left and right branches of the boundary of the stability region of conical states. This means that 
modulated states and the FFM state do not coexist in any region of the $(h,\Gamma)$ plane.

\subsection{Outstanding features of the stability diagram \label{sec:features}}

There are some characteristics in the stability diagram which have interesting consequences both from a theoretical and applied point of view. Below we enumerate these remarkable features and some of their consequences. Of especial relevance is the discussion of point 3 below.

 
\textit{1. Destabilization of $p$-states}.
At each point $(h,\Gamma)$ within the stability region of conical states the stable $p$-states 
are those whose stability ellipse encloses the point. 
Since all stability ellipses are centered at the origin in the $(h,\Gamma)$ plane, the only stable $p$-states are those which are metastable at $h=0$ and $\Gamma=0$, that is, which are metastable in absence of applied field and current. They are precisely those with $p$ in the range \eqref{eq:pbounds}.
The application of a field and/or a current does not stabilize any other $p$-state, but it
destabilizes some of them. As the point 
$(h,\Gamma)$ moves away from the origin, it crosses some ellipses, and the corresponding $p$-states become unstable. Outside the stability region of conical states, whose boundary is given by eqs.~\eqref{eq:envelope}, no $p$-state is stable.

\textit{2. Range of stable $p$-states}.
One conclusion of the discussion of point 1 above is that at each point $(h,\Gamma)$ the stable $p$-states have $p$ in a certain range 
$\pmin(h,\Gamma) \leq p \leq \pmax(h,\Gamma)$.
These two values, $\pmin(h,\Gamma)$ and $\pmax(h,\Gamma)$, are given by the two real roots of
\begin{equation}
\label{eq:ell2}
  A(p)\Gamma^2 + 2B(p)h\Gamma + C(p)h^2-D(p) = 0,
\end{equation}
that lie within the bounds given by Eq.~\eqref{eq:pbounds}. These two values, $\pmin$ and $\pmax$, approach each other as $(h,\Gamma)$ attains the stability boundary of conical states.
Therefore, the closest $(h,\Gamma)$ is to this stability boundary, the narrower the range 
of $p$ values of stable $p$-states.

\textit{3. Manipulating conical states}.
The discussion of point 1 above suggests a method to switch between metastable conical 
states with different wave number.
For instance, suppose we start at $h=0$ and $\Gamma=0$ with some metastable $p$-state, say 
with $p\approx 1$. 
If we apply a field and a polarized current such that $(h,\Gamma)$ corresponds to a point 
close to one of the red points of Fig.~\ref{fig:phd}, the initial $p$-state becomes unstable and it 
will evolve to one of the $p$-states within the stability range at $(h,\Gamma)$. Since this point 
is close to one of the red points of Fig.~\ref{fig:phd}, where the stability range is narrow, the final $p$-state will have $p\approx 2$. Since this state is metastable also for $h=0$ and 
$\Gamma=0$, it will remain as the field and the current are switched off. Hence, the consequence 
of this process is to switch the $p$-state from $p\approx 1$ to $p\approx 2$.
In Sec.~\ref{sec:numerics} we will show using numerical simulations that these processes are feasible.

Therefore, a given $p$-state can be selected with high precision by approaching the appropriate point of the stability boundary. The values of $h$ and $\Gamma$ appropriate to select a conical state with wave number 
$q \approx p q_0$ are those represented in Fig.~\ref{fig:pbound}.

\textit{4. Helicity switching}.
Since for the easy-plane anisotropy considered in this work $\hc>1$, we have that 
$1-\sqrt{\hc}<0$. This means that there are $p<0$ within the stability range \eqref{eq:pbounds}, and 
the corresponding $p$-states are stable within their stability ellipse.
These $p$-states with $p<0$ are conical states with helicity against the DMI. 
The metastability of these states opens the possibility of helicity switching in monoaxial chiral helimagnets through the action of a polarized current, by means of the process described in point 3 above.

\textit{5. Ferromagnetic states}.
For the same reason $p=0$ is within the bounds \eqref{eq:pbounds}. Hence, the $p$-state with $p=0$ is metastable within its stability ellipse (the black ellipse in Fig.~\ref{fig:phd}).
These states are ferromagnetic, with a uniform magnetization which has a component  
$n_z = \cos \theta_p = h/(\hc-1)$ along the chiral axis. The magnetization component lying on the plane perpendicular to the chiral axis is undetermined, which means that these ferromagnetic states are highly degenerate. 
This degeneracy is tantamount to the translational degeneracy of the conical $p$-states. 
Notice that these ferromagnetic states are different from the FFM state obtained for a sufficiently large magnetic field, in which the magnetization is aligned with the field.
Instead, they would be the equilibrium states in absence of DMI, which survive as metastable states  when the DMI is present.

\textit{6. Supercritical modulated states}.
The fact that the stability region of the FFM state is convex (see Fig,~\ref{fig:phd}) implies that there are steady moving $p$ states for supercritical applied fields ($|h|>\hc$). This means that if we start with the FFM state with $h$ in an appropriate range, such that $|h|>\hc$, and apply a polarized current of appropriate intensity the FFM state will be destabilized and will evolve to attain a steady moving $p$-state: a modulation will be created by the polarized current at a supercritical fileld. If the current is switched off, the FFM state will be recovered. One process of this kind is illustrated by micromagnetic simulations in Sec. \ref{sec:numerics}.

\section{Manipulation of the conical states}
\label{sec:numerics}

\begin{figure}[t!]
\centering
\includegraphics[width=1.0\linewidth]{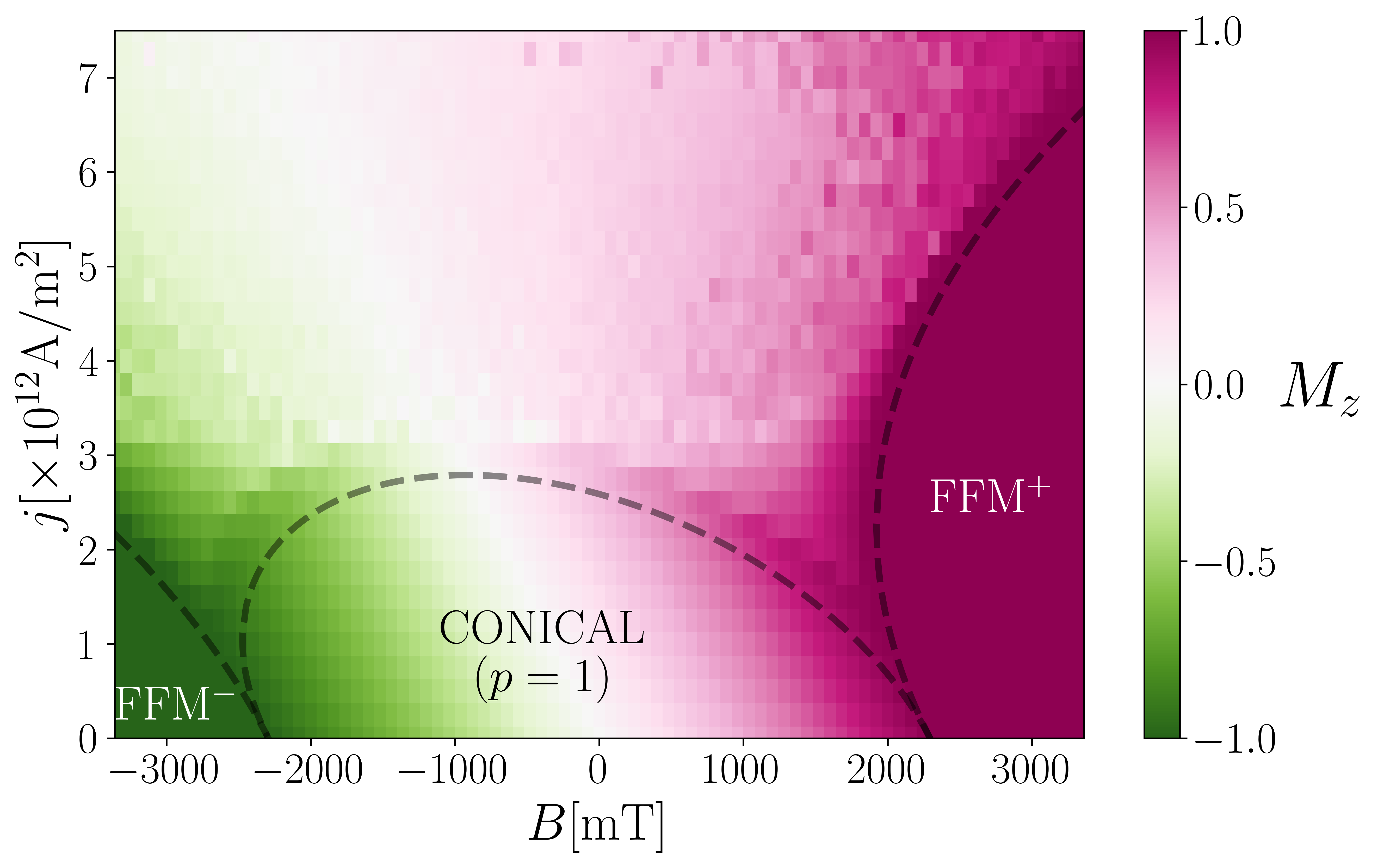}
\caption{Stability region for the equilibrium $p=1$ state ($q = q_0$) for applied magnetic field $B$ and current $j$, as obtained by numerical simulations. The regions where the FFM states are stable are also shown. The dashed lines correspond to the stability limits.
\label{fig:num1}}
\end{figure}


The peculiarities of the stability diagram suggest a method to manipulate the conical states described in Sec.~\ref{sec:features}, point 3. In this section we illustrate  that these ideas are sound by solving the LLG equation with appropriate initial conditions and time dependent applied field and current, by means of micromagnetic simulations.

The  micromagnetic numerical simulations are performed using the MuMax3 code~\cite{MuMax3,Leliaert18} in which a monoaxial DMI was implemented~\cite{Laliena20} to model the system given by Eqs.~\eqref{eq:E} and \eqref{eq:torque}, with material parameters appropriate for \CrNbS~\cite{Laliena20}: $A = 1.42$~pJ/m, $D = 369$~$\mu$J/m$^2$, $K = -124$~kJ/m$^3$, and $M_\mathrm{S} = 129$~kA/m (see Osorio \textit{et al.}~\cite{Osorio2021} for further details). With these parameters the equilibrium pitch of the helical state is $L_0 = 2 \pi/q_0 \approx 50$~nm and the critical magnetic field $\hc$ corresponds to $B_c \approx 2300$~mT. The simulations are performed for a one-dimensional system of linear size $R = 500$~nm, with a mesh size  $\Delta R = 1$~nm, and periodic boundary conditions. We set $\alpha = 0.01$ and $\beta = 0.02$. Notice that in a finite system with periodic boundary condition only a discrete number of $2 \pi$ rotations can be attained. We denote by $Q$ the winding number (see Ref. ~\onlinecite{Osorio2021} for the definition of $Q$), and $Q_0 = R/L_0 = 10$ is the equilibrium winding number, which corresponds to the $p$-state with $p = 1$. Hence $p = L_0/L = Q/Q_0$ takes only discrete values with step size $1/Q_0=1/10$.

\begin{figure}[t!]
\centering
\includegraphics[width=1.0\linewidth]{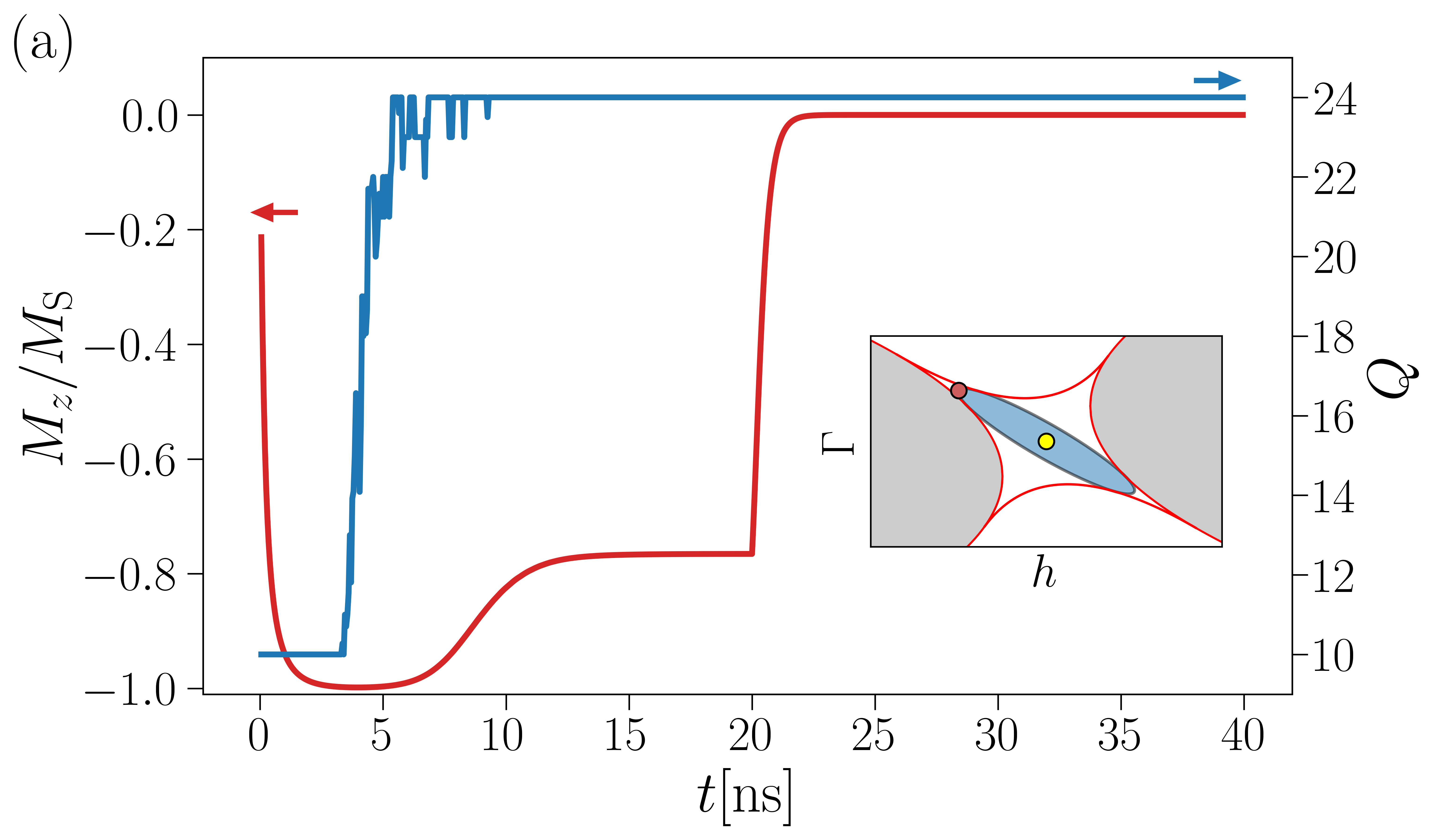}
\includegraphics[width=1.0\linewidth]{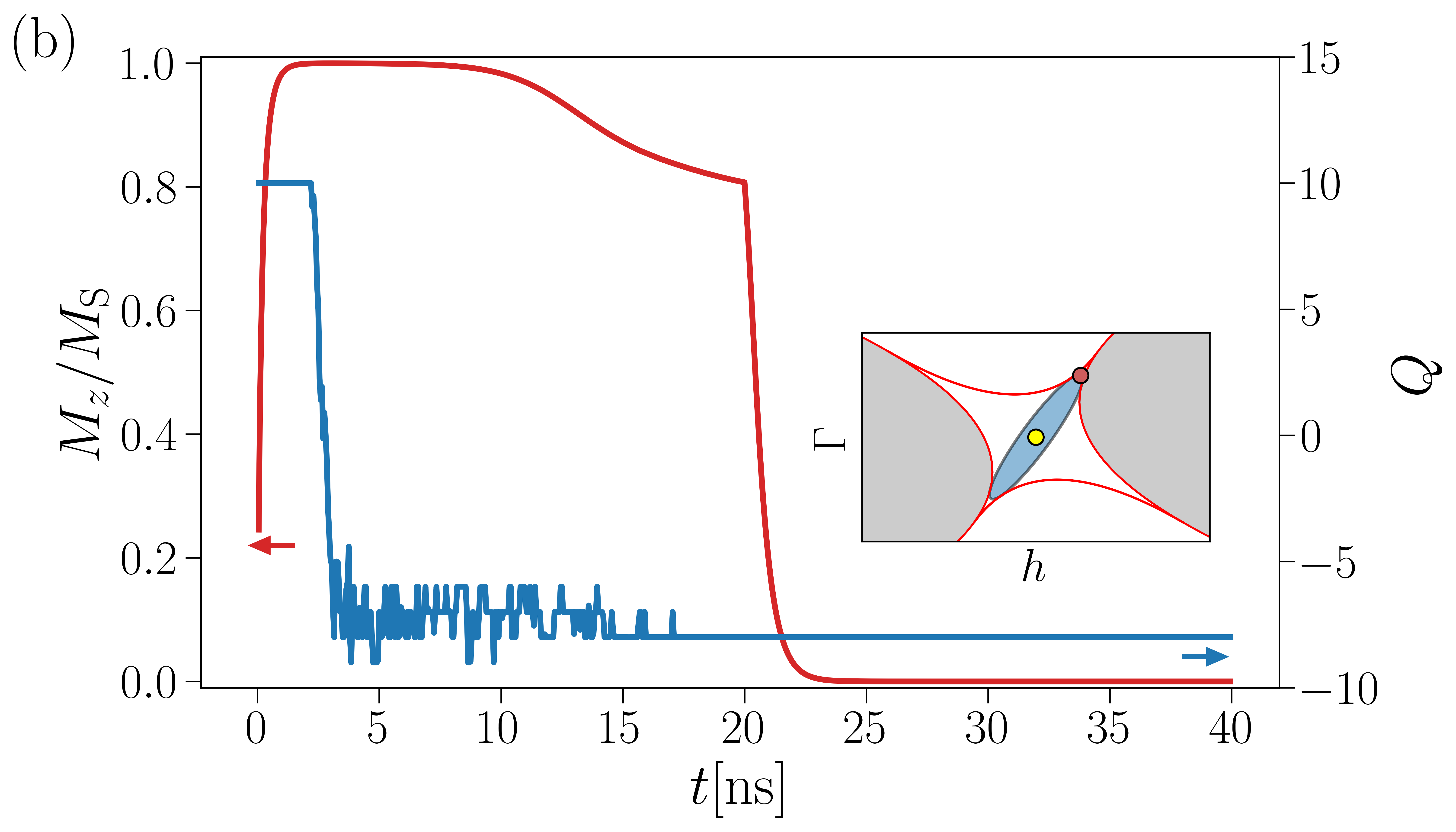}
\caption{Evolution of the system when a simultaneous $(h_a,\Gamma_a)$ square pulse is applied during 20 ns: the red and blue curves represent the net magnetization along $z$ and the winding number, respectively. Initially the system is at $(h = 0, \Gamma = 0)$ (yellow point in the inset) and is characterized by a winding number $Q_0=10$. The values of $(h_a,\Gamma_a)$ (red points in the inset) are in one semi-axis of the shown ellipses corresponding to positive $p = 2$ in (a) and negative $p = -0.5$ in (b).}
\label{fig:num2}
\end{figure}

Let us start showing the stability diagram corresponding to the equilibrium $p$-state ($p=1$) obtained from numerical simulations. 
Given a point $(h,\Gamma)$ of the stability diagram, the initial state is either the $p$-state with $p=1$ if $|h+\Gamma|\leq \hc$ (this is the existence condition of the $p=1$ state), or the FFM state, otherwise. A perturbation of small intensity, $M_{\mathrm{S}}/10$, and random orientation is added to the magnetization of the initial state. Then, the current is applied for 20 ns. Figure~\ref{fig:num1} displays the stability region and the corresponding stability ellipse (see. Fig.~\ref{fig:phd}). The stability limits of the FFM are also shown in the figure. The stability boundaries are in good agreement with the analysis in Sec.~\ref{sec:Stability-FFM}.

In order to show how the system can be manipulated to obtain different targeted $p$-states we use the following protocol. The system is initialized at $(h=0,\Gamma = 0)$ in a state with a winding number $Q_0 = 10$, which corresponds to the equilibrium state with $p = 1$. A small random perturbation is added to the three components of the magnetization, so that the initial state is actually a perturbed $p$-state.
Then a simultaneous square pulse of magnetic field and polarized current with $(h=h_a,\Gamma = \Gamma_a)$ is applied during $20$~ns, afterwards the system is let to relax to some metastable state at $(h=0,\Gamma = 0)$. The evolution of the system can be followed by monitoring the time evolution of the magnetization along the chiral axis, which accounts for the conical distortion, and the winding number $Q \propto p$.
A small pulse length is necessary to destabilize the initial state. We expect the final state to depend on the initial perturbation and the shape of the pulse, but not on the length of the pulse, since ones a new state has been reached it is metastably retained.

Figure~\ref{fig:num2} shows how different $p$-states can be stabilized following the main diagonals of the ellipses. In Fig.~\ref{fig:num2}(a) results are presented for $(h_a < 0,\Gamma_a >0)$, while results with $(h_a > 0,\Gamma_a >0)$ are shown in Fig.~\ref{fig:num2}(b). In both cases, $(h_a,\Gamma_a)$ lies outside the stability region of the $Q_0$ state.
In Fig.~\ref{fig:num2}(a) a final state with $Q = 24$ ($p = 2.4$) is obtained, showing that a combined pulse of magnetic field and current can be used to modify the winding number ($p$-state) in the system. In (b), the values of $(h_a > 0,\Gamma_a >0)$ are such that the only stable $p$-states are those with $p<0$. Numerical results show that in this case a helicity switching can be forced with the final state metastably retained against the DMI favored rotation direction.
This shows how the magnetic configuration can be destabilized in favor of new $p$-states by pushing the system outside the ellipse corresponding to the initial state.

\begin{figure}[t!]
\centering
\includegraphics[width=1.0\linewidth]{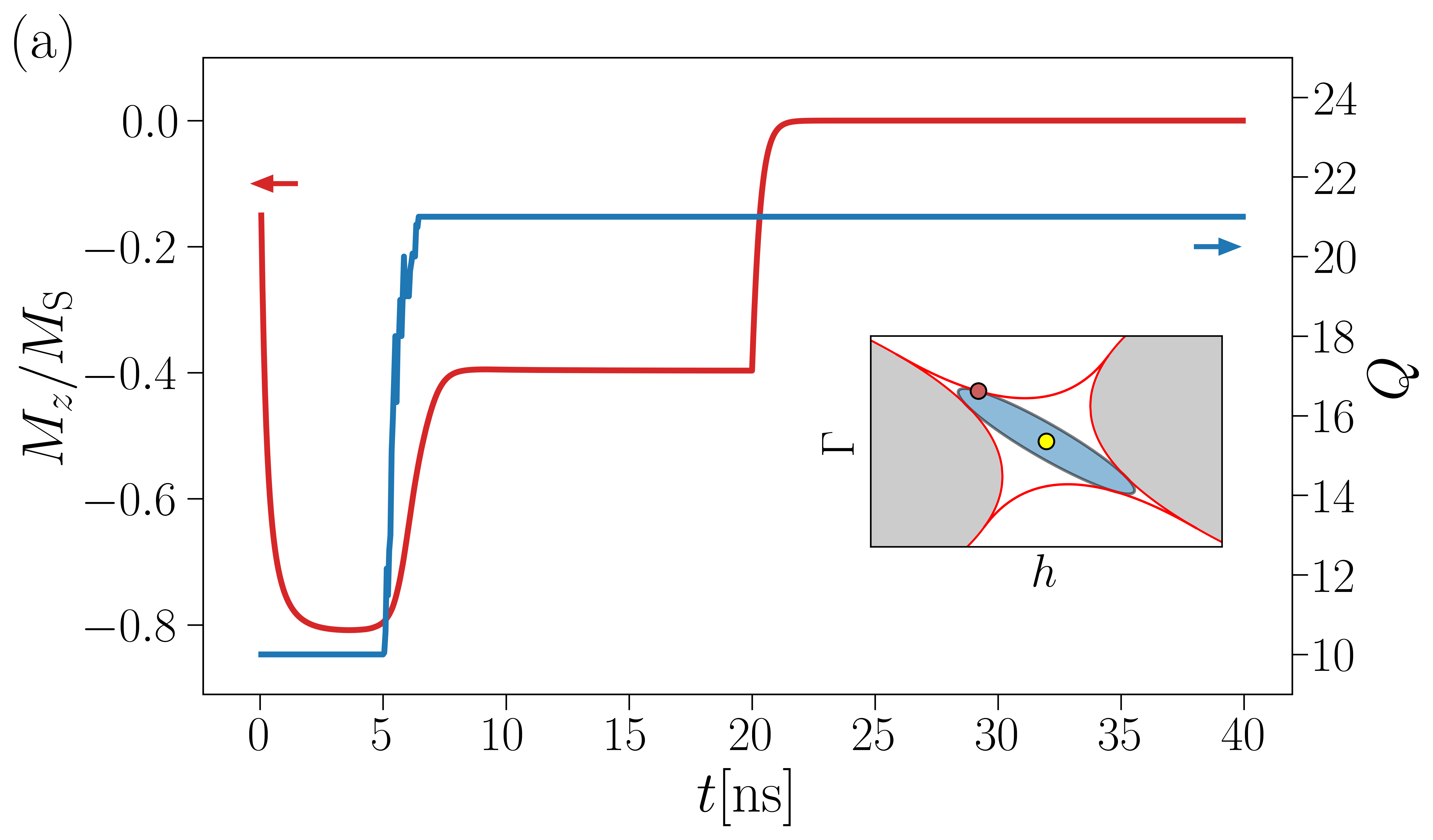}
\includegraphics[width=1.0\linewidth]{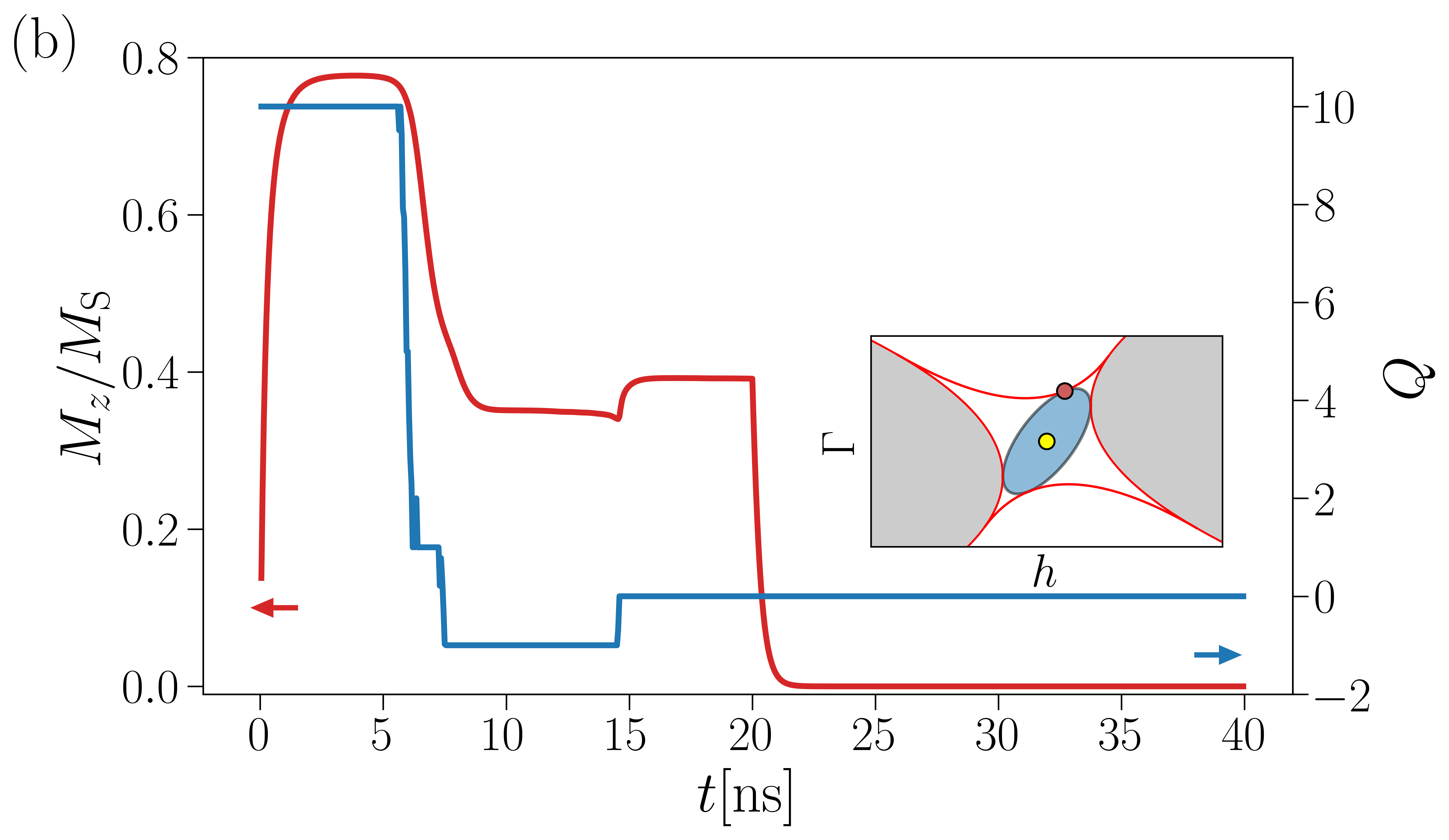}
\caption{Evolution of the system when a simultaneous $(h_a,\Gamma_a)$ square pulse is applied during 20 ns: the red and blue curves represent the net magnetization along $z$ and the winding number, respectively. Initially the system is at $(h = 0, \Gamma = 0)$ (yellow point in the inset) and is characterized by a winding number $Q_0=10$. The values of $(h_a,\Gamma_a)$ (red points in the inset) are exactly at the point where the ellipses for (a) $p = 2$ and (b) $p = 0$ touch the upper branch of the stability boundary, as shown in the insets.
\label{fig:num3}}
\end{figure}

In order to select a $p$-state with a targeted $p$ value, $h_a$ and $\Gamma_a$ should be chosen at the stability boundaries, such that $\pmin \leq p \leq \pmax$ with $\pmin$ and $\pmax$ very close to each other. Figure~\ref{fig:num3} present results using $(h_a,\Gamma_a)$ close to the stability boundaries and such that (a) $p = 2$ and (b) $p = 0$ are expected (see square symbols in Fig.~\ref{fig:pbound}). Numerical results show that these targeted $p$-states can be readily selected. In Fig.~\ref{fig:num3}(a) $Q = 22$ is obtained, which corresponds to $p = 2.2$, close to the targeted $p=2$ state. Since we use $(h_a, \Gamma_a)$ exactly at the point where the ellipse for $p=2$ touches the stability boundary, the $p=2$ state is in its stability limit and states very close to $Q=20$ can be stabilized, in this case $Q=22$. In Fig.~\ref{fig:num3}(b), after fluctuating around $Q = \pm 1$, the final ferromagnetic state with $Q = 0$ is obtained. Note that this state is initially (when $h=h_a$) oriented along a random direction within the cone with $n_z = \cos \theta_p = h_a/(\hc -1)$, depending on the initial perturbation of the system. When $h=0$ the obtained ferromagnetic state is contained in the easy-plane ($xy$) defined by the magnetic anisotropy. This ferromagnetic state can be metastably retained, as opposed to the FFM state.

\begin{figure}[t!]
\centering
\includegraphics[width=1.0\linewidth]{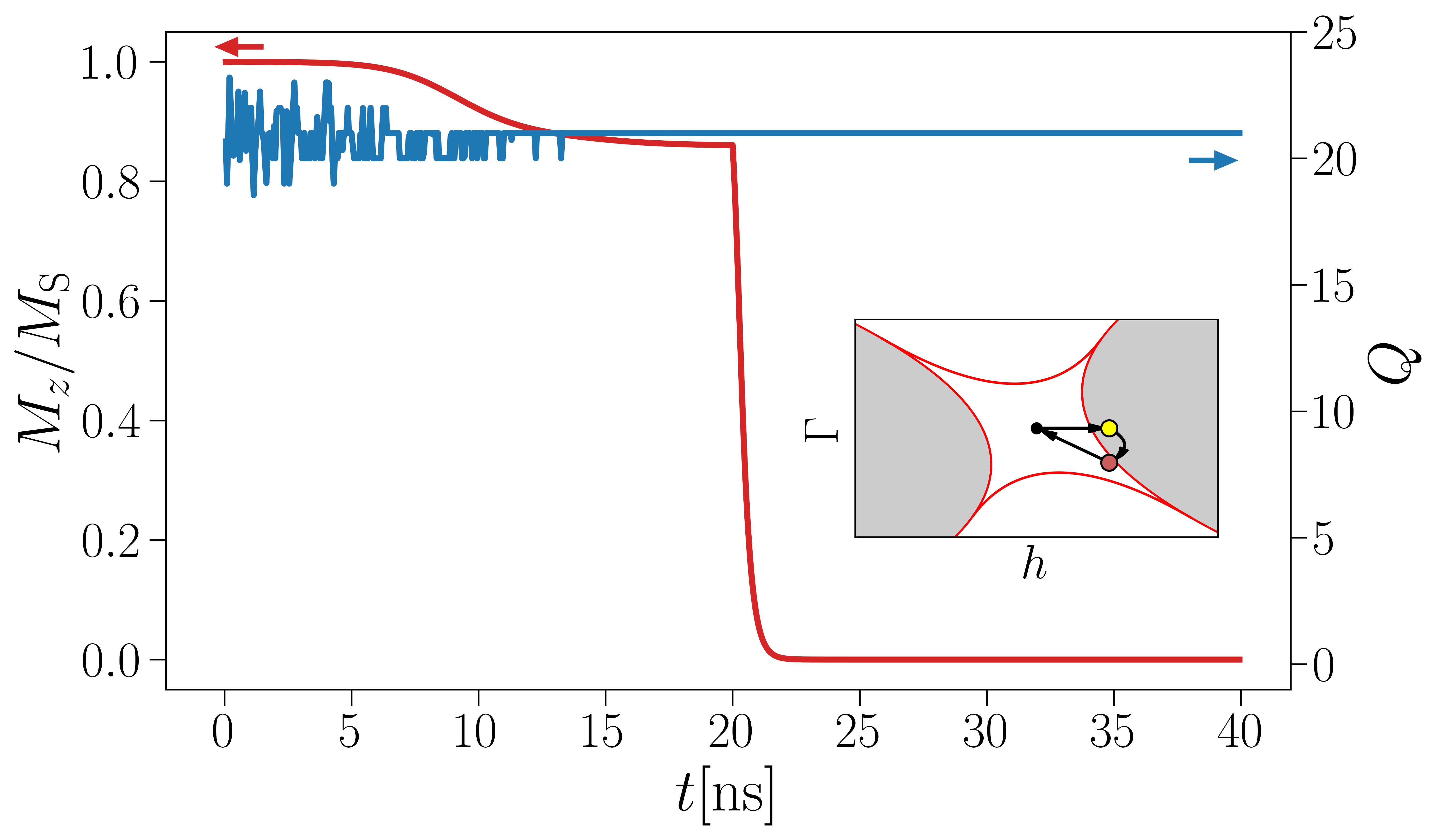}
\caption{Switching to a metastable state at zero field using independent field and current pulses. First, a field pulse of intensity $h>\hc$ is used to drive the system to the FFM state (yellow point in the inset). Then, a $t=0$ a current pulse is applied using a negative $\Gamma$ value (red dot in the inset) to stabilize the $p$-state. After 20 ns the system is let to relax to a metastable state (the obtained $p$-state) at ($h = 0,\Gamma = 0$).
The red and blue curves represent the net magnetization along $z$ and the winding number, respectively.
\label{fig:num4}}
\end{figure}

For $\Gamma = 0$, going beyond $\hc$ erases the $p$-state and the FFM state is stabilized. 
It is important to note that in this case, for a field value $h>\hc$, an applied current can be used to stabilize a $p$-state, as shown in Fig.~\ref{fig:num4}. The system is initialized with $Q_0 =10$ at $(h=0, \Gamma=0)$ and then is set in a FFM state using $(h=h_a>\hc, \Gamma=0)$. Applying then $\Gamma = \Gamma_a$ inside the stability region, a state with $Q = 21$ is obtained. This state remains when going back to $(h=0, \Gamma=0)$. 
That is, some $p$-states can be created by means of a two step process: first, the current $p$-state is erased by applying a field higher than $\hc$ and afterwards the FFM state is destabilized by applying an appropriate current. The system evolves to some steady moving stable $p$-state which remains after the field and the current are switched off.

The numerical results shown in this section illustrates how the stability diagram obtained in Sec.~\ref{sec:Stability-conical} can be used to manipulate the conical states.

\section{Conclusions}

Let us summarize the findings reported in this work.
Besides the equilibrium state, at low temperature and zero applied field monoaxial chiral helimagnets have 
a continuum of helical states differing by the wave number of the modulation \cite{Laliena18b}, which 
can be written as $pq_0$, where $q_0$ is the wave number of the equilibrium state and $p$ is a dimensionless number. These states are called here the $p$-states. For an infinite system, their energy is a continuum function of $p$ which is minimized by the equilibrium state, corresponding to $p=1$. These states 
are local minima of the energy for $p$ in a neighborhood of $p=1$ \cite{Laliena18b}. We argued 
here (Sec. \ref{sec:conical}) that, in spite of what the curve energy \textit{versus} $p$ may suggest (Fig.~\ref{fig:eph}), the $p$-states are metastable in that range. 

The application of a magnetic field parallel to the chiral axis has two effects: first, it introduces a conical deformation of the $p$-states; and second, it schrinks the interval of metastability. For applied fields of strength higher than the critical field no $p$-state is stable, and the equilibrium state is the FFM state.

Analogously, the application of a polarized current along the chiral axis has three effects on the $p$-states: first, they reach a steady moving state with a velocity proportional to the intensity of the applied current; second, they suffer a conical deformation similar to that introduced by the application of an external field in the direction of the chiral axis; and third, some $p$-states are destabilized, and therefore the stability interval is schrinked.

The most remarkable fact of the stability diagram of $p$-states in the applied magnetic field - applied current intensity plane (Sec. \ref{sec:features}) is that for each $p$ in the stability range at zero field there are points in the stability diagram at which the interval of stability is very narrow and contains such $p$. This fact allows us to devise processes to select a given $p$-state. For instance, if we start with some metastable $p$-state at zero current and apply appropriate magnetic field and current we end with a steady moving $p$-state with wave number within a narrow interval around the targeted $p$. These new $p$-state is metastable at zero field and zero current and therefore it would remain as the field and the current are switched off. The feasibility of these processes, which is 
 extremely important from the point of view of applications, is shown by micromagnetic simulations (Sec. \ref{sec:numerics}).
 
Switching between $p$-states opens the possibility of their application in spintronic devices. In particular, there are metastable $p$-states with negative $p$, and therefore helicity switching is possible in monoaxial chiral helimagnets.
It has been argued that controlled switching among magnetic states with opposite helicity might be used for memory applications~\cite{Jiang2020}. Current induced helicity switching has been discussed before in a non-chiral monoaxial helimagnet~\cite{Ohe2021} and in isolated skyrmions in frustrated magnetic systems~\cite{Zhang2023}. In both cases magnetic textures of pure exchange origin were studied, while we report here helicity switching in a monoaxial \textit{chiral} helimagnet, where chirality is due to the presence of DMI. 

The $p$-states exist also in cubic chiral magnets \cite{Laliena17b}. In that case they are characterized not only by the wave number, but also by the orientation of the wave vector. The dynamics and stability of the $p$-states of cubic chiral helimagnets under the action of a polarized current have been recently studied by Masell \textit{et al.} \cite{Masell2020combing,Masell2020manipulating} in the zero applied field case. These authors showed that, under the action of the current, the $p$-states reach a steady motion state with a velocity proportional to the current density and, at the same time, they suffer a uniform conical deformation with an angle determined by the current. This behaviour is the same found here for monoaxial chiral helimagnets. Masell \textit{et al.} found also a critical current which destabilizes the $p$-states. Their analysis of the longitudinal Fourier modes, whose wave vector is parallel to the $p$-state wave vector, gives a destabilizing current which exactly coincides with the result reported here for monoaxial helimagneets, in the particular case $h=0$, $\hc=1$, as it must be since by ignoring the transverse fluctuations the cubic chiral helimagnet becomes the monoaxial chiral helimagnet without single-ion anisotropy (UMA). 
In addition, they found that the $p$-states are destabilized by any current, however small, applied perpendicularly to the wave vector of the $p$-state. This means that the $p$-states tend to propagate along the direction of the applied current. The situation becomes more interesting if there is also an applied field, since in this case the propagation direction of the $p$-state tends to be aligned with the field. Thus, the interplay between the magnitude and relative orientation of the applied field, the applied current, and the $p$-state wave vector promises a complex and rich stability diagram of $p$-states in cubic chiral helimagnets.

The essential question of the lifetime of metastable $p$-states cannot be addressed with the methods of this work. The $p$-states are separated in the magnetic configuration space by energy barriers (they are local minima of the energy functional) and their lifetime depends on the height of such barriers. 
Thus, it is clear that the lifetime will increase by decreasing the temperature and, therefore, the presence of metastable $p$-states will be more easily detected at low temperature.

The above discussion on lifetimes is related to the experimental signals of the $p$-states. To address these questions it is necessary a careful analysis of the experimental data at low temperature to seek for anomalies attributable to $p$-states. We have already remarked that $p$-states exist also in cubic chiral helimagnets \cite{Laliena17b}.
In these systems the continuum of $p$-states is richer than in monoaxial chiral helimagnets since, besides the wave number, the $p$-states differ also in the orientation of their wave vectors. 
The low temperature anomalies reported recently for the cubic chiral helimagnet MnSi \cite{Ohkuma22} may be due to the presence of metastable $p$-states.





\begin{acknowledgments}
Grant Number PGC2023XM4 funded by MCIN/AEI/10.13039/501100011033 supported this work. Grants OTR02223-SpINS from CSIC/MICIN and DGA/M4 from Diputaci\'on General de Arag\'on (Spain) are also acknowledged.
This work was also supported by the Grant No. PICT 2017-0906 from the Agencia Nacional de Promoción Científica y Tecnológica, Argentina.
\end{acknowledgments}

\appendix

\section{Stability of the conical states}
\label{app:conical}

A necessary condition for the stability of a $p$-state is that the spectrum of the $2\times 2$ matrix operator $\mathcal{D}$, whose matrix elements are the linear differential operators given by Eqs.~(\ref{eq:D11p})-(\ref{eq:D22p}) lies in the complex half-plane with non positive real part. 
The $p$ dependence is hidden in the parameters $a$, $\Delta$, and $b$ given by Eqs. (\ref{eq:ap}) and (\ref{eq:dbp}). Since the coefficients of those linear operators are constants, the spectrum is easily obtained by Fourier transform.

Denoting the wave vector of the Fourier modes by $\vk$, and setting  $k=|\vk|$, the spectrum is given by two complex functions of $\vk$, denoted by $\lambda_{\pm}(\vk)$, given by
\begin{equation}
\begin{split}
\lambda_{\pm}(\vk) =& -\frac{\alpha}{2}(2k^2+a) \\[6pt]
&+\iu\big(\Delta-(1+\alpha\beta)b\big)k_z
\pm \sqrt{a_r+\iu\,a_i}.
\label{eq:spectrum}
\end{split}
\end{equation}
where
\begin{gather}
  a_r = \frac{\alpha^2a^2}{4}-k^2(k^2+a) + \big(\alpha\Delta + (\beta-\alpha)b\big)^2k_z^2, \\[6pt]
  a_i = -\big(\alpha\Delta+(\beta-\alpha)b\big)(2k^2+a)k_z.
\end{gather}
We only need the real parts, which are given by
\begin{equation}
\mathrm{Re}\,\lambda_{\pm}(\vk) = -\frac{\alpha}{2}(2k^2+a) \pm \sqrt{\frac{\sqrt{a_r^2+a_i^2}+a_r}{2}},
\end{equation}
Since $\mathrm{Re}\lambda_+(\vk) \geq \mathrm{Re}\lambda_-(\vk)$, stability requires 
$\mathrm{Re}\lambda_+(\vk) \leq 0$, which, with simple algebraic manipulation, it can be shown to be equivalent to
\begin{equation}
\big(\alpha\Delta + (\beta-\alpha)b\big)^2k_z^2\leq \alpha^2 k^2(k^2+a). \label{eq:ineq2}
\end{equation}
Since the left hand side of this inequality is non negative, and since it must hold for any $k$, and in particular for $k\to0$, we get $a\geq 0$. This relation sets the bounds for the $p$ values of stable $p$-states given by the inequalities (\ref{eq:pbounds}).

Since $a\geq 0$, the right hand side of (\ref{eq:ineq2}) increases with $k_x^2+k_y^2$, and therefore the inequality is satisfied if and only if it is satisfied for $k_x^2+k_y^2=0$. Thus we set $k^2=k_z^2$, and then we have
\begin{equation}
k_z^2\left(k_z^2+a - (\Delta+q_0\Gamma)^2\right) \geq 0,
\end{equation}
Noticing that with $a \geq 0$ inequality (\ref{eq:ineq2}) holds for all $\vk$ if and only if it holds for $\vk=k_z\hz$, the stability condition reduces to
$(\Delta+q_0\Gamma)^2 \leq a$. 
To arrive to this inequality Eq. (\ref{eq:jGamma}) was used. 
Substituting the expressions for $a$, $\Delta$ and $\cos \theta_p$ in 
this inequality we get the following expression for the stability condition:
\begin{equation}
A(p)\Gamma^2+2B(p)\Gamma h+C(p) h^2 \leq D(p), \label{eq:ellipse_app}
\end{equation}
where
\begin{eqnarray}
&&A(p) = 2(p-1)^3+3(\hc+1)(p-1)^2+ \nonumber \\
&& \qquad \qquad 6 \hc(p-1)+\hc(\hc+1), \\[6pt]
&&B(p) = (p-1)^3+3(p-1)^2+3\hc(p-1)+\hc, \hspace*{1cm} \\[6pt]
&&C(p) = \hc+3(p-1)^2, \\[6pt]
&&D(p) = \big(\hc-(p-1)^2\big)^3. \label{eq:Dp}
\end{eqnarray}
Notice that $D(p)\geq 0$ for $p$ satisfying the bounds (\ref{eq:pbounds}). 
The inequality (\ref{eq:ellipse_app}) determines a region in the $(h, \Gamma)$ plane limited by a conic section. The discriminant of the left hand side of (\ref{eq:ellipse_app}) is
\begin{equation}
B(p)^2 -A(p)C(p) = -D(p) \leq 0.
\end{equation}
Therefore, the conic section is actually an ellipse centered at $(0, 0)$ with the principal axes rotated with respect to the coordinate axes. The amount of rotation depends on $p$. The steady moving $p$-state is stable within the region of the $(h,\Gamma)$ plane enclosed by the corresponding ellipse. 
The stability the static $p$-states discussed in Sec.~\ref{sec:conical} is obtained as a particular case of this general approach, setting $\Gamma=0$. The static $p$-state is thus stable 
in the range of $h$ determined by the intersection of its stability ellipse with the $\Gamma=0$ axis.

The region of the $(h,\Gamma)$ plane in which there exists some stable steady moving $p$-state is
bounded by the envelope of the one-parametric family of ellipses given by Eq. (\ref{eq:ellipse}).
The envelope can be readily found and it has four branches determined by the parametric equations
\begin{eqnarray}
  & &\left\{
  \begin{array}{l}
    h=-\big[ (p-1)^2+2 (p-1)+\hc\big] \\[4pt]
    \Gamma = 2(p-1)
  \end{array}
  \right. \label{eq:env1} \\[6pt]
  & &\left\{
  \begin{array}{l}
    h=(p-1)^2+2 (p-1)+\hc \\[4pt]
    \Gamma = -2(p-1)
  \end{array}
  \right. \label{eq:env2} \\[6pt]
  & &\left\{
  \begin{array}{l}
    h=-\left[(p-1)^2+2 \hc (p-1)+\hc \right]/\sqrt{\hc} \hspace*{1cm}\\[4pt]
    \Gamma = \left[(p-1)^2+\hc\right]/\sqrt{\hc}
  \end{array}
  \right. \label{eq:env3}\\[6pt]
  & &\left\{
  \begin{array}{l}
    h=\left[(p-1)^2+2 \hc (p-1)+\hc \right]/\sqrt{\hc} \\[4pt]
    \Gamma = -\left[ (p-1)^2+\hc \right]/\sqrt{\hc}
  \end{array}
  \right. \label{eq:env4}
\end{eqnarray}
with $p$ is in the range given by Eq.~\eqref{eq:pbounds}.

The parameter $p$ can be eliminated in each of these four pairs of equations and then 
the equations of the envelope in the form (\ref{eq:envelope}) are obtained. 
This envelope bounds the region of the 
$(h,\Gamma)$ plane where some (steady moving) $p$-state is stable, which in Sec. \ref{sec:Stability-conical} is called the \textit{stability region of conical states}.



\section{Stability of the FFM}
\label{app:FFM}

To linear order, the dynamics of perturbations, $\xi$, of the FFM state obey Eq.  (\ref{eq:LLGlin2}), in this case with
\begin{eqnarray}
  &&\mathcal{D}_{11} =\alpha(\nabla^2-a)-\big(2q_0+(1+\alpha\beta)b\big)\partial_z, \\[6pt]
  &&\mathcal{D}_{12} = \nabla^2-a+\big(\alpha 2q_0-(\beta-\alpha)b\big)\partial_z, \\[6pt]
  &&\mathcal{D}_{21} =-\nabla^2+a-\big(\alpha 2q_0-(\beta-\alpha)b\big)\partial_z,\\[6pt]
  &&\mathcal{D}_{22} = \alpha(\nabla^2-a)-\big(2q_0+(1+\alpha\beta)b\big)\partial_z,
\end{eqnarray}
where now
\begin{equation}
  a=q_0^2(|h|+\kappa), \quad b = \sigma(h)\frac{q_0^2b_jj}{\omega_0}=q_0\frac{\alpha}{\beta-\alpha}\sigma(h)\Gamma,
  \label{eq:ab2}
\end{equation}
with $\sigma(h)=1$ if $h\geq 0$ and $\sigma(h)=-1$ if $h<0$.

If the FFM is stable the spectrum of $\mathcal{D}$ lies on the complex half-plane with non positive real part.
Again, the spectrum of $\mathcal{D}$ is easily obtained by Fourier transform. If, as before, $\vk$ is the wave vector of the Fourier mode, the spectrum is given
by the complex functions $\lambda_{\pm}(\vk)$, whose real parts are
\begin{equation}
    \mathrm{Re}\,\lambda_{\pm}(\vk) = -\alpha(k^2+a) \pm \big(\alpha2q_0-(\beta-\alpha)b\big)k_z.
\end{equation}
Now, $\mathrm{Re}\,\lambda_{\pm}(\vk)\leq 0$ if and only if
\begin{equation}
\alpha k_z^2 \pm \big(\alpha2q_0-(\beta-\alpha)b\big)k_z + \alpha a \geq 0,
\end{equation}
for all real $k_z$. This means that the two roots in $k_z$ of the left hand side of the above inequality must be either complex or equal, that is, the discriminant of the quadratic polynomial in $k_z$ given by the left hand side of the above inequality must be non positive:
\begin{equation}
\big(\alpha2q_0-(\beta-\alpha)b\big)^2 - 4\alpha^2 a \leq 0.
\end{equation}
Inserting the values of $a$ and $b$ given by equation (\ref{eq:ab2}) and defining $\Gamma$ by 
Eq. (\ref{eq:jGamma}) we obtain
\begin{equation}
\Gamma^2-4\sigma(h)\Gamma+4(\hc-|h|)\leq 0.
\end{equation}
To have a non-empty solution of this inequality the two roots in $\Gamma$ of its left hand side must be real, and then the inequality holds for $\Gamma$ being between the two roots.
Then we get the condition $|h|>\hc-1$ and, if this holds, the two roots are given by
\begin{equation}
\Gamma_{\pm}=\sigma(h)2\big(1 \pm \zeta(h) \big),
\end{equation}
with $\zeta(h) = \sqrt{1+|h|-\hc}$.
In this way we obtain that the stability region of the FFM state in the $(h,\Gamma)$ plane is determined by the inequalities (\ref{eq:FFMstability}). It is remarkable that the boundary of the stability region of the FFM state coincides exactly with two of the branches of the boundary of the stability region of conical states. As stressed at the end of Sec. \ref{sec:Stability-FFM}, this means that conical states never coexist with the FFM state.

\bibliography{references}

\end{document}